# Neutrons production on the IPHI accelerator for the validation of the design of the compact neutron source SONATE.

A. Menelle[1], F. Ott[1], F. Prunes[1], B. Homatter[1], B. Annighöfer[1], F. Porcher[1], C. Alba-Simionesco[1]

[1] Laboratoire Léon Brillouin, CEA, CNRS, Université Paris-Saclay, CEA Saclay 91191 Gif sur Yvette France

N. Chauvin[2], J. Schwindling[2]

[2] IRFU/SACM, Université Paris-Saclay, CEA Saclay 91191 Gif sur Yvette France

A. Letourneau[3], A. Marchix[3], N.H. Tran[3]

[3] IRFU/SPhN, CEA, CNRS, Université Paris-Saclay, CEA Saclay 91191 Gif sur Yvette France

## 1 Abstract

We aim at building an accelerator based compact neutron source which would provide a thermal neutron flux on the order of $4 \times 10^{12}$ n.s$^{-1}$.cm$^{-2}$.sr$^{-1}$. Such brilliance would put compact neutron sources on par with existing medium flux neutron research reactors. We performed the first neutron production tests on the IPHI proton accelerator at Saclay. The neutron fluxes were measured using gold foil activation and 3He detectors. The measured fluxes were compared with MCNP and GEANT4 Monte Carlo simulations in which the whole experimental setup was modelled. There is a good agreement between the experimental measurements and the Monte-Carlo simulations. The available modelling tools will allow us to optimize the whole Target Moderator Reflector assembly together with the neutron scattering spectrometer geometries.

## 2 Introduction

There is currently an interest in developing compact neutron sources CNS based on low energy proton accelerators (10-100 MeV) [1]. Such sources could serve as neutron sources for Boron Neutron Capture Therapy [2-3] or as neutron for neutron scattering to replace small ageing nuclear reactors [4]. There are already several projects of CNS on-going around the world. The currently most advanced is LENS Low Energy Neutron Source at Indiana University [5]. The CNS operating or under construction have gathered into the UCANS, Union for Compact Accelerator-driven Neutron Sources [6].

At Saclay we are considering a similar possibility to replace the Orphée research reactor. Our first aim has been to experimentally validate the neutron production and moderation obtained by Monte Carlo simulations using either MCNP [7] or GEANT4 [8]. Once reliable simulation tools are available we shall be able to estimate the performances of a CNS for neutron scattering experiments (from the source to the spectrometer) and compare its performances to existing facilities (reactor or spallation based).

As a starting point we have used the IPHI proton accelerator which shall eventually be able to produce high current proton beams (up to 100 mA CW) with proton energies of 3 MeV. Our final goal is to eventually boost the proton energy to 20 MeV.

# 3  Experimental setup

## 3.1  Proton source

The experiments have been performed on the IPHI proton accelerator which is based at CEA Saclay, France [9]. The accelerator consists of a proton source SILHI of energy 95 keV, a Low Energy Beam Transport Line and a Radio Frequency Quadrupole to accelerate the protons to an energy of 3 MeV. The accelerator is designed to operate in continuous mode with proton currents up to 100 mA which corresponds to a total power of 300 kW. For the current experiments dedicated to validations of neutron production and moderator Monte Carlo simulations we have operated the accelerator at a very low power of about 10 W, both to avoid any target damage and for radioprotection issues. The accelerator was operated in pulsed mode with proton pulses of length 100 µs and with a repetitoin rate of 1 Hz and a peak current of 30 mA. Rather narrow thermal neutron pulses are thus obtained so that precise time-of-flight measurements can be performed.

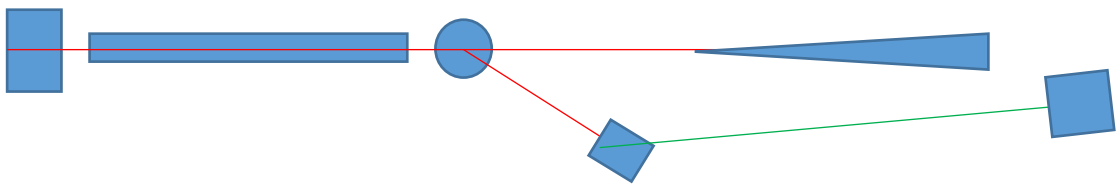

*Figure 1: Sketch of the experimental set-up. The target was installed on a deviator line and the neutron detectors were set at 8 meters from the target so as to be able to perform time-of-flight measurements.*

In order to produce neutrons we opted for a beryllium target of thickness 0.5mm (99.0%) which stopped all incident protons. The target was attached on an aluminum support with titanium screws in order to minimize neutron activation. This support was air cooled. The proton beam size was limited to 16mm in diameter. The proton current incident on the target was continuously measured so as to have a precise value of the incident particle flux and be able to precisely estimate the neutron production. An electron repelling electrode set at a potential of 200 V was set in front of the beryllium target so as to prevent any bias in the target current measurement. The target was installed in a polyethylene (PE) moderator box (300x300x400 mm$^3$) so as to cool down the neutron to thermal energies (around 26meV). A 20 mm diameter hole was drilled through the moderator from the position where the thermal neutron density was expected to be the highest to the outside of the PE box (see Figure 2a).

In order to change the moderator geometry, it was possible to insert PE plugs inside the exit hole so as to fill more or less the neutron extraction channel.

The whole experimental setup (accelerator – target – moderator – detectors) was installed in a 2 m thick concrete casemate.

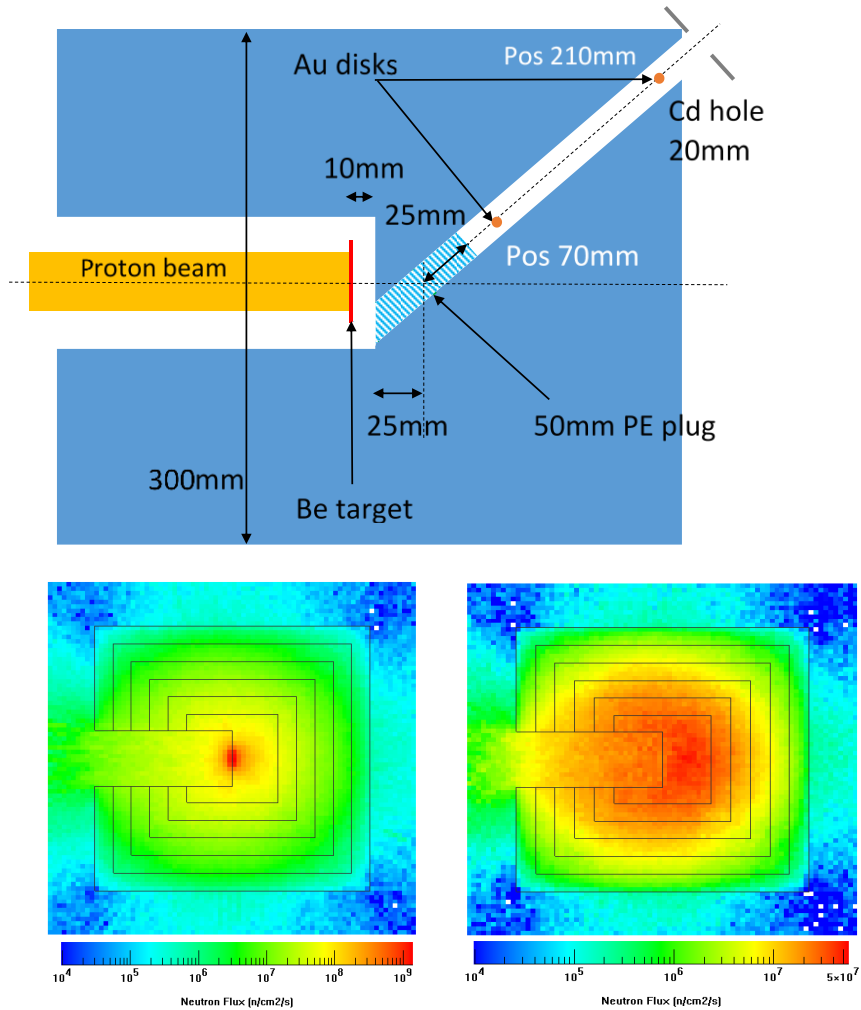

Figure 2: *(a) Target and moderator assembly; (b) fast neutron densities; (c) thermal neutron densities (E<100meV) in n/cm²/s/µA.*

## 3.2 Neutron detection

Several diagnostic tools were installed. In order to monitor the fast neutron production (via the proton incident on the *Be* target and/or accelerator elements), a Bonner sphere was installed in the casemate. We set-it up as close as possible from the thermal neutron detectors (at 8.4 m from the target). A set of five vertical $^3$He tubes (diameter 50mm, length 130 mm was installed in a PE box (20 mm thick PE + an extra $B_4C$ elastobore lining) but without any front collimation. The large array of detectors did not provide quantitative results because the lack of front collimation led to the measurement of "spurious neutrons" moderated in the casemate walls. Besides since we were operating in pulsed mode with rather short pulses, the neutron flux was very close or above the neutron detector saturation (20 kHz). Thus in order to provide reliable quantitative results, a single horizontal detector $^3$He was used (diameter 50mm, length 130mm) and set at 8.4m from the target in a heavier shielding box providing some front collimation (angular opening of about 10°). The $^3$He pressure in the tube was 5 bars so that over the tube length of 130mm, the detection efficiency is 100% down to 0.9 Å. It slightly drops to a value of 95% at 0.6 Å and 65% at 0.2 Å. Hence it can be considered that the thermal neutron spectrum does not need to be corrected for detection efficiency. One may be worried by the long length of the detector which may bias the ToF measurements. If we aim for a time resolution of 1%, we need to have the neutrons detected in the first 1% x 8.4m = 84mm of the tube. For neutron wavelengths above 0.7 Å, 90% of the particles are detected in these first 84 mm of the $^3$He tube. Hence we can assume that the ToF is measured with an accuracy better than 1% for thermal neutrons.

This detector was covering a solid angle of 0.3x0.3° so that the instantaneous neutron flux was significantly reduced. A neutron sensitive image plate was also used but is essentially provided information about the gamma fluence.

Being able to quantify all types of secondary particles produced during the stripping reaction (fast neutrons, gamma radiations and thermal neutrons) is important in order to be able to design properly shielded neutron scattering instruments. Two gamma sensitive detectors were thus also installed around the moderator.

Thermal neutrons (1.8 Å) travel at a speed of 2200m/s so that the travel time over 8.4 m is 3.8 ms. For a proton pulse width of 100 µs, the energy resolution is thus on the order 0.1/3.8 = 2.6% which is good enough to measure a rather precise neutron energy distribution. Note that besides the proton pulse length, the moderation process leads to an intrinsic neutron pulse broadening which reduces the energy resolution. Monte-Carlo simulations provide an estimate of a broadening on the order of 100-150 µs in a PE moderator [10-11].

In order to validate the Monte-Carlo simulations inside the moderators, gold disk were positioned at various positions inside the moderator. Disks with a diameter 6 mm and thickness 200 µm were used. The weight of these disks was in the 100 mg range so that even when operating at 10W, measurable activation of the gold was achieved within 15-30 minutes of operation. Bare disks and disks enclosed in Cadmium were measured so as to be able to estimate the thermal and fast neutron flux [10]. The Cd absorption cut-off edge is at around 500 meV (σ = 100 barn) so that the bare disk are activated by the whole neutron spectrum irradiation and the Cd covered disk are only activated by the fast neutrons. These gold activation measurements provide a simple and quantitative way of measuring the thermal neutron flux. However, while the measurement of the gold activation is quantitative and reliable, in order to calculate an absolute neutron flux, an a priori knowledge of the neutron energy distribution is necessary. Hence either an assumption of neutron energy distribution has to be made (Maxwellian distribution centered around 26meV for example) or the activation must be calculated using a simulated neutron energy distribution. Gold disks (with and without Cd casing) have been put into the moderator model and their activation has been calculated during the Monte Carlo simulations. The measured activation values have then been compared with the calculated activation values.

# 4 Experimental results

## 4.1 Gold-foil activation

Gold disks were activated at various positions inside the moderator (at 70mm and at 210mm from the maximum thermal flux position). A third measurement was performed with a very small moderator of diameter (66mm) and thickness (50mm) set at the back of the target. The gold disks were irradiated for durations ranging from 15 to 22 minutes which led to measurable gamma activity from $^{198}$Au (except for the fast neutron measurement, with Cd, at 210mm) which was below the detection limit.

It is possible to have a first estimate of the neutron flux by assuming that inside the moderator we have a thermal Maxwell Boltzmann distribution. By making this assumption it is possible to estimate the neutron flux corresponding to the measured activity. A neutron activation calculator [12] tells us that a neutron fluence of $10^8$ n/cm² gives rise to an activation of 90 Bq/g of $^{198}$Au.

|  | Measured activity (Bq/g) | Proton fluence (mC) | Measured activity (Bq/g/µC) | Calculated activity (Bq/g/µC) | Neutron flux assuming thermal MB (n/cm²/µC) |
|---|---|---|---|---|---|
| Position 1 (70mm, 15min) | | | | | |
| Without Cd | 2700 | 2.6 | 1.04 | | 1.2 x10$^6$ |
| With Cd | 420 | 2.6 | 0.162 | | - |
| Position 2 (210mm, 23min) | | | | | |
| Without Cd | 57 | 3.6 | 0.016 | | 1.8x10$^4$ |
| With Cd | <9.7 | 3.6 | <0.003 | | - |
| Small Moderator (15min) | | | | | |
| Without Cd | 140 | 2.9 | 0.048 | | 5.4 x10$^4$ |
| With Cd | 140 | 2.9 | 0.048 | | - |

*Table 1: Gold disk activations for various experimental conditions. The measured activities can be compared to the calculated activation using GEANT4.*

Previous measurements using a very similar moderator geometry were performed at a proton energy of 10 MeV and at proton currents of 30 µA [13]. The neutron brilliance at a distance of 70 mm from the moderator axis, also measured by the gold foil method was 3x10$^9$ n/cm²/s (Figure 3 in Allen et al). It is possible to rescale our results by applying a neutron yield gain factor of 48 between proton energies of 3 MeV to 10 MeV [4] and a factor 30 to renormalize per µC. The renormalized flux (from 10 to 3 MeV) would thus be 2x10$^6$ n/cm²/µC. Both values are consistent, especially considering the fact that the neutron energies and the moderator geometries are somewhat different.

## 4.2 Time-of-flight measurements

The time-of-flight measurements were performed by setting a single $^3$He detector at a distance of 8.4m in the direction of the extraction hole. The incident neutron flux was measured with time channels of width 20 µs. The start of the acquisition was synchronized with the clock signal of the accelerator radio-frequency. Each ToF spectra was acquired in 5 to 10 minutes. Figure 3 shows examples of ToF spectra measured in various moderator geometry. The neutron signal was measured with the bare target (without any PE moderator around it, dark blue signal). A small moderator ("Mini-moderator") consisting of a block of diameter 66mm and thickness 50mm was set right after the target. Then the neutron signal was measured with the PE box around the target and several filling of the extraction hole (hole totally empty, filled with a 50mm long rod and filled with a 130mm rod). For these last 3 measurements, the moderator box was encased in a B$_4$C shielding so as to measure only the neutrons exiting the moderator from the 20mm diameter Cd hole (see Figure 2a). The first time channels (~100 µs) correspond to epithermal neutrons with energy above 1 eV. Thermal neutrons (~26 meV, 1.8 Å) take 3.8 ms to travel over the 8.4 m.

In the case of the measurements performed with the PE moderator box, one observes quite a lot of epithermal neutrons (below 1ms), the more so as the presented data are not corrected for detector efficiency. This is followed by a thermalized neutron peak which has a maximum at around 2.5ms, corresponding to a neutron wavelength of 1.2A°. The maximum of the distribution correspond to a Maxwellian distribution centered around a temperature of 350K (see Figure 3c). The neutron spectra are very close in the case of an empty hole or with a 50 mm plug. In the case of the empty hole, the distribution is a bit wider which might be accounted for by the fact that the emission surface is ill define in this case. On the other hand, when the extraction hole is totally filled there are very few neutrons emitted by the moderator. One may wonder why so many epithermal neutron are observed. This can be rather simply explained by the fact that around the moderator, the B$_4$C shielding is only stopping thermal neutrons while all the epithermal and fast neutrons are still emitted from the whole moderator volume and not only from the 20 mm diameter hole as the thermal neutrons. Hence, the epithermal and thermal flux cannot be quantitatively compared. The measurement without any moderator and with a small moderator (green and dark blue) show of course significantly more epithermal neutrons. They also show a hump around the thermal peak at 2ms which can be accounted by moderation in some parts of the accelerator. The third striking point is that a long tail of slow neutrons (t>4ms) is clearly visible on Figure 3b in log scale. This does not reflect the presence of long wavelength neutrons but rather the fact that fast neutrons are moderated in the concrete of the casemate and are then travelling back as thermal neutrons into the detectors. These various effects make these measurements very difficult to exploit quantitatively.

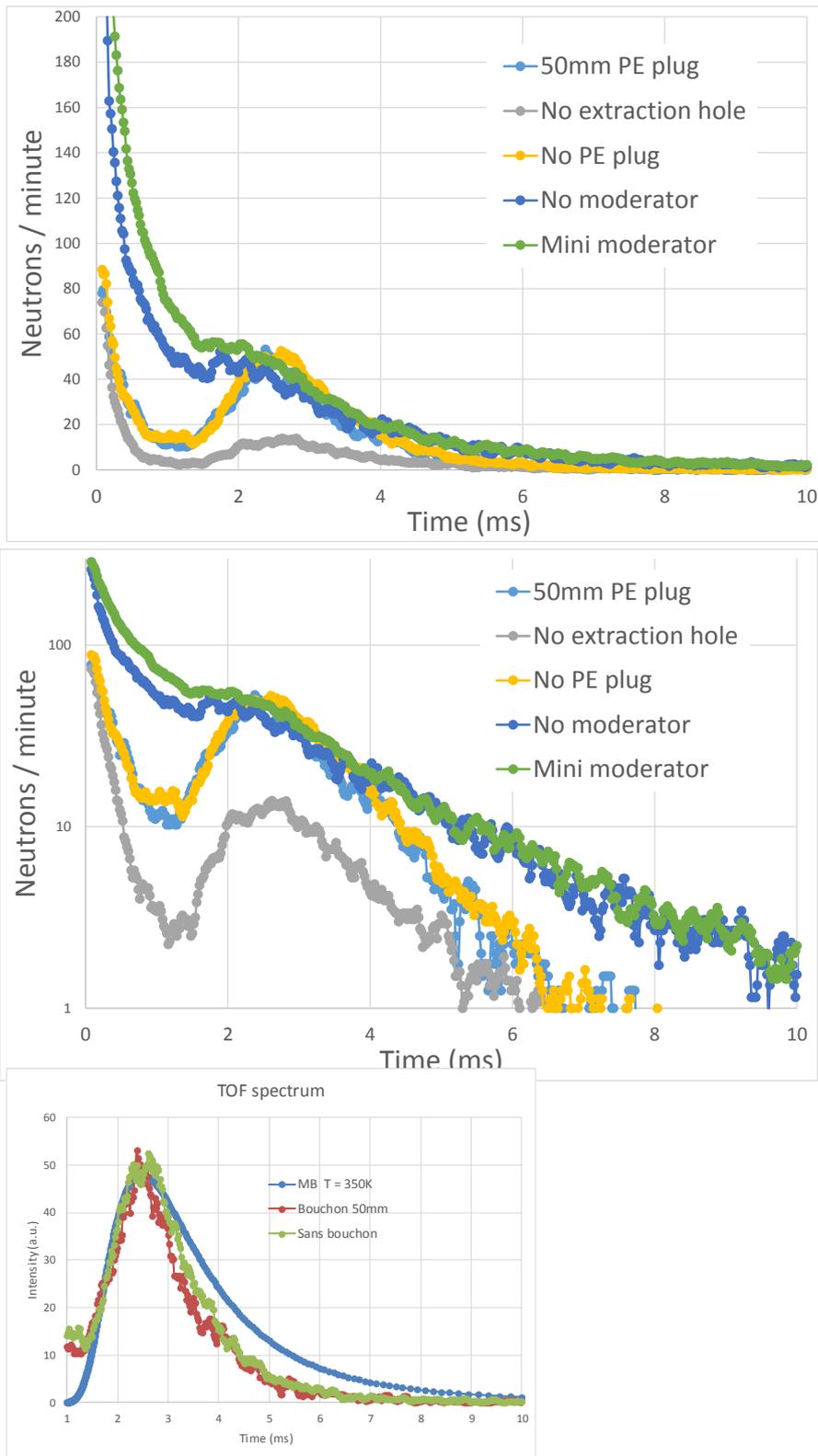

Figure 3: *Time-of-flight spectra measured for various moderator geometries. Time channels of 20 μs have been used. (a) linear scale. (b) log scale. (c) fit of the thermal signals with a Maxwell Boltzmann distribution with T= 350K.*

## 4.3 Diffraction experiment

In order to have a measurement of the neutron pulse shape and of the time-of-flight parameters, we used a graphite single crystal. The LENS team is using a Ge crystal which provides better peak shapes [5]. In the transmission geometry, crystal at 0° and detector in the direct beam, the results were not satisfactory due to a high background level on the detector. Hence, we performed the measurement in a diffraction geometry with the crystal set at 45° with respect to the incident neutron beam. The crystal was installed at 8.4 meters from the source and the detector was set at 90° from the incident neutron beam. Hence it was possible to perform a Time-of-flight diffraction experiment on the (00l) graphite diffraction peaks. Due to the 90° position of the detector, the background noise was significantly reduced compared to the transmission geometry. The incident spectrum was first measured in the direct beam without crystal. The diffraction spectrum was then measured with the detector at 90° and at a distance of 30 cm from the crystal. The diffracted spectrum was divided by the incident spectrum to normalize the neutron intensities.

(a)

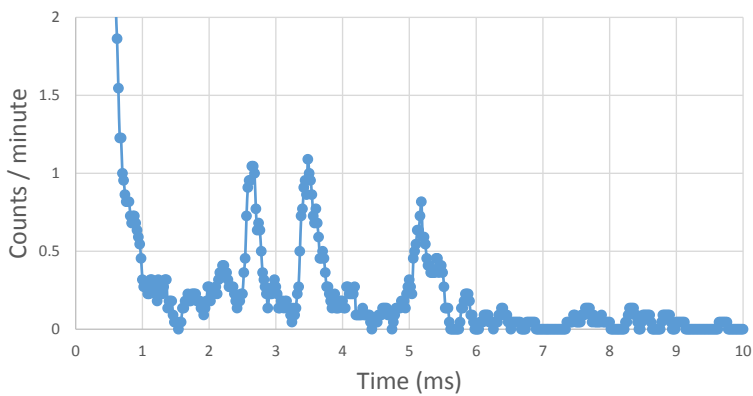

(b)

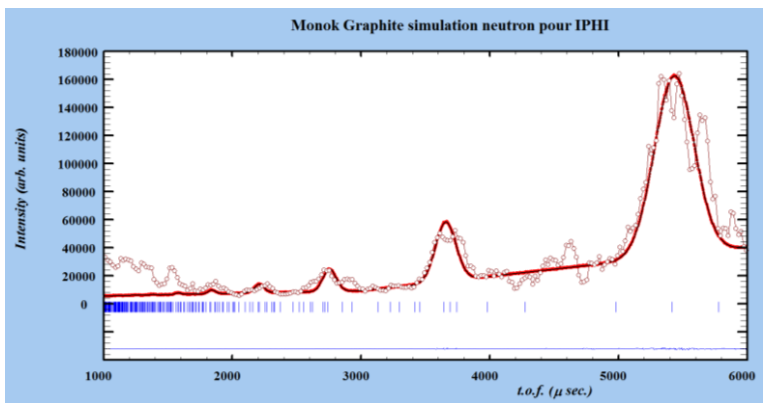

*Figure 4: (a) Raw time-of-flight spectra measured on a graphite single crystal. (b) Normalized diffraction spectrum and fit using FullProf. The relative measured intensities are satisfactory.*

## 4.4 Gamma production

The production of gamma radiation was followed during the whole experiments. Without any moderator, the measured gamma activity was 70 µSv/h at a distance of 2m from the target. With the PE moderator set in place and without any dedicated gamma radiation shielding, the gamma background was measured at around 50 µSv/h at 2 m from the source.

Monte-Carlo simulation (MCNP or GIANT4) provide the gamma radiation spectrum generated by the Target-Moderator assembly.

| Run | mC | Balise 1 LDO | Balise 2 LDO | Balise 1 LDO.h/µC | Balise 2 LDO.h/µC | Balise 1 Sv/µC | Balise 2 Sv/µC | Dose 1 µSv/h | Dose 2 µSv/h |
|---|---|---|---|---|---|---|---|---|---|
| 1 (29 Juin 11h35) 15min | 2.6 | 1.03 | 1.96 | $10^{-4}$ | $1.9\times10^{-4}$ | $2.5\times10^{-9}$ | $4.75\times10^{-9}$ | 26 | 49 |
| 2 (4 Juillet 15h05) 23min | 3.6 | 0.94 | 1.81 | $10^{-4}$ | $1.9\times10^{-4}$ | $2.5\times10^{-9}$ | $4.75\times10^{-9}$ | 23.5 | 45 |
| 3 (5 Juillet 15h15) 15min | 2.9 | 1.85 | 2.78 | $1.6\times10^{-4}$ | $2.4\times10^{-4}$ | $4\times10^{-9}$ | $6\times10^{-9}$ | 46 | 70 |

*Table 2: Gamma radiation measurements.*

Image plate measurements with lead absorbers allowed to estimate that at 2 meters from the moderator assembly, the peak gamma spectrum energy was in the 100 keV range.

# 5 Discussion

GEANT4 was used to perform Monte-Carlo simulations of a Target-Moderator assembly to calculate the neutron flux produced in a stripping reaction between protons at 3 MeV and beryllium. We have shown that very low power (~10 W) is sufficient to perform a characterization of the source. Nevertheless higher power (~100W) would be more comfortable to perform diffraction experiments.

From these results it is possible to make a first extrapolation of the neutron brilliance which could be achieved by scaling the proton energy from 3MeV to 20MeV and the proton current from 1 µA to 100 mA. A gain in the neutron yield of a factor 200 can be achieved by increasing the proton energy from 3 MeV to 20 MeV. The brilliance at the surface of the moderator would be $1.2\times10^6$ n/cm²/µA/s x $10^5$ µA x 200 = $2.4\times10^{13}$ n/cm²/s. This brilliance is close to the brilliance of the Orphée reactor at the entrance of the guide systems which is on the order of $1.5\times10^{14}$ n/cm²/s. It is likely that an optimized moderator geometry could rise the brilliance on par with that of the Orphée reactor. However, such an accelerator based source could not operate in continuous mode due to the huge heat load on the target (20 MeV x 100 mA = 2 MW). Such a source should be operated in pulsed mode with a typical duty cycle on the order of 2-4 %. This would correspond to a power load on the target on the order of 40-80 kW which is way more manageable. A detailed account on the possibilities of such a source to perform neutron scattering experiments will be published elsewhere [14].

# 6 Conclusion and outlook

The IPHI accelerator was operated for the first time to produce thermal neutrons. Further experiments will be performed to test cold moderator geometries. Besides, the issue of a beryllium target subject to an intense proton current is a delicate engineering issue. At least a dozen groups are working on this issue, either to produce neutron for physics experiments or for Boron Neutron Capture Therapy. It is thus very likely that solutions will be found in the short term to be able to operate such sources at power in the 100kW range. Hence compact neutron source would probably demonstrate performances such that they can replace small nuclear research reactors for a wide range of experiments ranging from radiography, neutron scattering, activation analysis…